# Removing fluid lensing effects from spatial images.

Greg Sabella

University of Adelaide

## 1. Introduction

Remote sensing of earths' land based ecosystems has been ongoing for many decades producing a variety of multispectral images at varying resolutions. This has been critical in allowing the study of ecosystems, global development, natural disasters and ultimately our changing planet. However, a comparable effort has not been made for aquatic ecosystems. Even though 100% of the lunar and Martian surfaces have been mapped at a spatial resolution of 100m or finer, as of 2018 only around 5% of Earth's seafloor has been mapped [1].

Shallow water and coastal aquatic ecosystems such as coral reefs and seagrass meadows play a critical role in regulating and understanding Earth's changing climate and biodiversity. They also play an important role in protecting towns and cities from erosion and storm surges. These ecosystems are highly sensitive to changes in land management, water management and climate. As such, monitoring the health and changes in these ecosystems is critical.

One example of the sensitivity of these ecosystems can be seen in the seagrass meadows along the coast of Adelaide in Australia [2]. Over the last 50 years around 1/3 of seagrass along the coast has been lost. The main cause of this loss has been from poor water quality resulting from drain discharge, storm water run-off and effluent disposal. Waves and currents further erode the exposed seafloor making the recovery of seagrass meadows difficult.

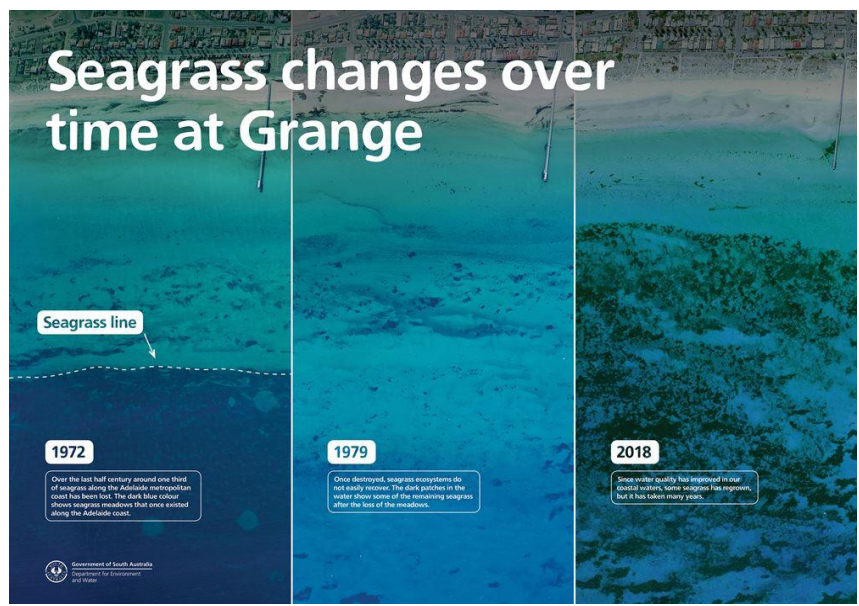

*Figure 1: Seagrass changes over time at Grange, Adelaide, Australia [2]*

Technologies used for remote sensing of aquatic systems from above the surface (drones, UAVs and satellites) cannot produce images with comparable effective spatial resolutions (ESR) to those over terrestrial areas. While commercial satellites can produce images with an ESR of 0.3m for terrestrial targets, producing images with an ESR finer than 10m over oceans results in a highly distorted image even on a calm day with the clearest of waters.

As stated by [1], the optical interaction of light with fluids such as oceans is a complex phenomenon. As visible light interacts with surface waves optical aberrations appear causing bands of light on the seafloor. The light is also attenuated through effects such as absorption and scattering. Refractive lensing, the magnification or demagnification of objects under the water as viewed from the surface, also occurs. These combined effects are known as the ocean wave fluid lensing phenomenon and is the reason why aquatic images taken above the surface are so distorted at lower ESRs.

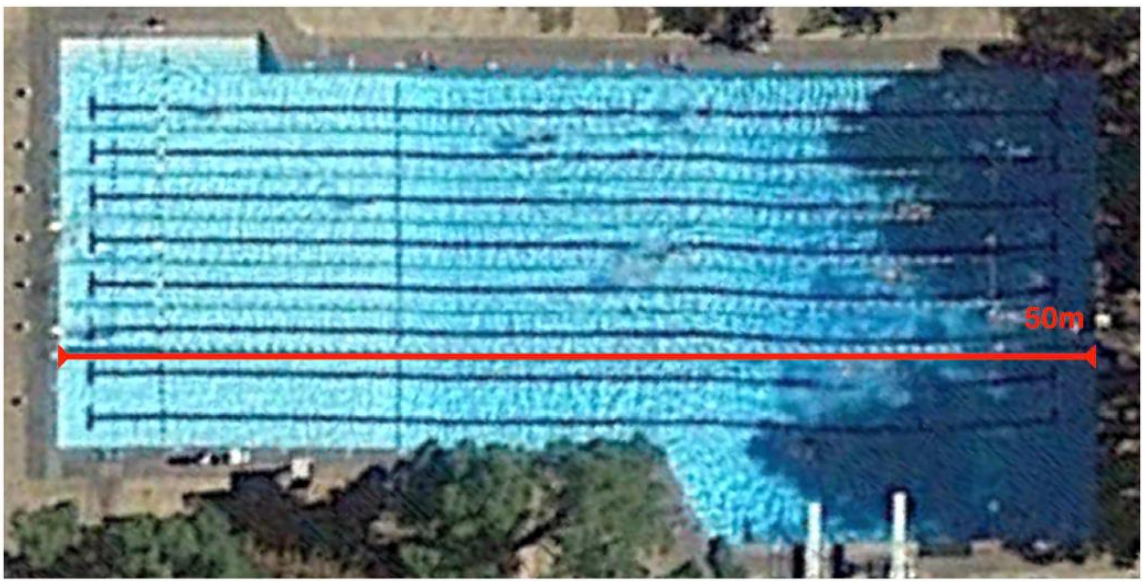

Figure 2: Satellite image of swimming pool that shows obvious impacts of fluid lensing by small surface waves and changes in depth (most notable in the pool lane lines) [1].

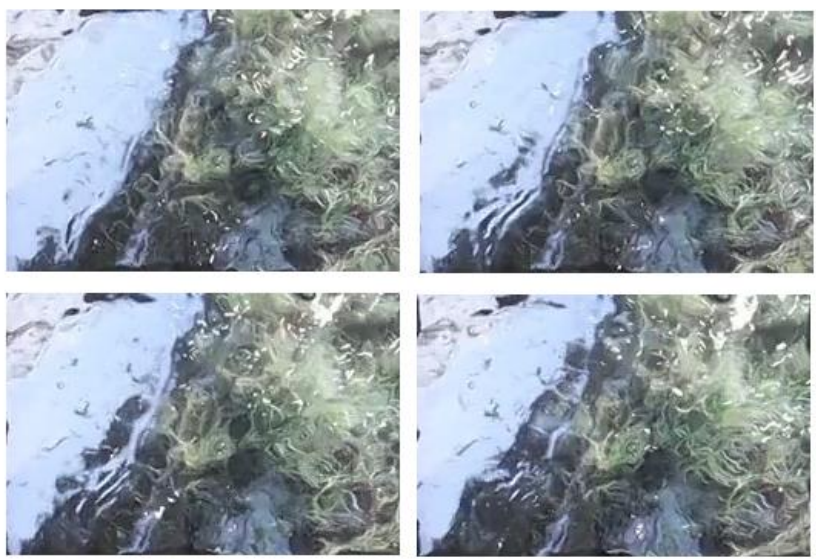

Figure 3: Video frames showing the changes in an object from the effects of fluid lensing.

As part of [1] [3] [4] NASA developed a fluid lensing algorithm to remove the effect of fluid lensing from images. This algorithm proved quite successful allowing photos to be taken of coral reefs with a much lower ESR. Unfortunately the algorithm code is not yet publicly available. There are no other known methods that effectively remove the fluid lensing effects from images.

The aim of this research is to see if machine learning can 'learn' the NASA algorithm allowing a more accessible model to be created that can be used to support the monitoring activities of the Adelaide seagrass restoration project. To test if this might be possible a simple proof of concept experiment was devised and has been the focus of work to date. A fluid lensing replication tank was set up to create datasets that could be used to train and evaluate models.

In this report a number of datasets are presented including a shapes dataset and an "outdoor seagrass" dataset. Several models are implemented and evaluated in their effectiveness at removing the effects of fluid lensing. These include super resolution based models, convolutional LSTMs, deblurring models and Spatial Temporal Convolutional (STCN) models. STCN combined with SIFT Flow for data pre-processing has proven to be the most effective model tested with results showing that it can remove most of the fluid lensing effects and recreate a clearer image of the underwater objects with only a slight loss of the finer details of the object.

## 2. Related Work

In 2016, NASA released results of an algorithm [1] [3] [4] that had been developed to remove the effects of fluid lensing from images. It a passive sensing method that is reliant on sunlight and is therefore limited to photic zone of the ocean – the first 100m in very clear water. The algorithm makes use of high frame rate video, temporal and positional data to perform image reconstruction. It consists of a number of stages.

Firstly the basic relationships between fluid lensing lenslets (peaks and troughs of surface waves), caustic cells (bands of refracted light), image formation, magnification and depth are determined. Secondly, the spatial locations of caustic focal planes, caustic cell size, lenslet periodicity, velocity and curvature are determined using a fluid distortion characterisation algorithm. The baseline median and mean image is also computed. Thirdly, a caustic bathymetry fluid lensing algorithm is used to uniquely characterize the bathymetry of the benthic scene from caustic behaviour alone. Fourthly, the derived features are used to constrain and precondition SIFT Flow and to produce a 2D reconstructed image.

This method has been validated through a number of field missions. The results look very impressive (see Figure 4). Using the fluid lensing algorithm, underwater objects are able to be seen in significantly greater clarity. Unfortunately, the algorithm code is not publicly available (it is patented) so the results of the research below could not be compared to this algorithm.

Other work has been done on related areas such as reconstructing fluid surfaces [5] however no other methods have been discovered so far that produce comparable results to [1].

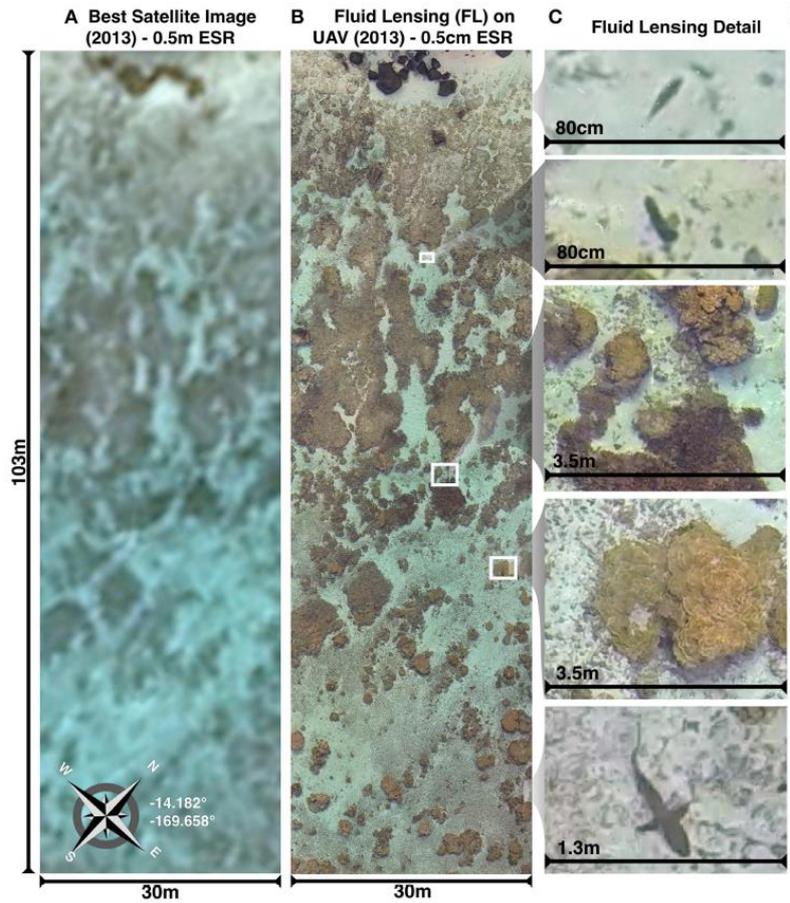

*Figure 4: (A) Highest-resolution publicly available image of a transect area captured June 2015 from Pleiades-1A satellite with 0.5m ESR. (B) Fluid lensing 2D result of the same area as captured from UAV at 23m altitude with estimated 0.5–3 cm ESR. (C) Inset details in fluid lensing 2D image include a parrotfish ~20 cm in length, a sea cucumber ~ 1 cm in length, multiple coral genera including Porites and Acropora, and a reef shark. [1]*

## 3. Methodology

To see if machine learning can remove fluid lensing effects from images using passively collected data, a simple proof of concept experiment was devised to test potential models. A fluid lensing replication tank was set up to create datasets that could be used to train and trial different machine learning methods. For the first dataset this was a plastic container with a static image stuck on the bottom. For the later datasets a tank made of transparent glass was used. This allowed a tablet to be placed underneath that could display different images (Figure 5). A camera was held in place above the tank so that images/videos would be aligned. The tank was filled with around 20cm of fresh water so that manually disturbing the water could reproduce the fluid lensing effects on the images being displayed. The tank was placed under a room light so that some light reflection was present on the water surface however this was not sufficient to reproduce the bands of light that appear on underwater objects as in Figure 2 and Figure 9.

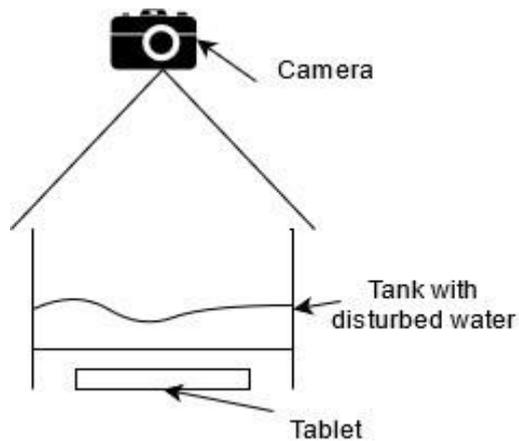 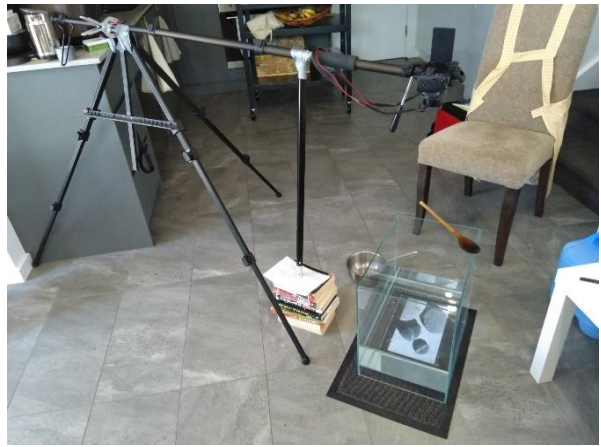

*Figure 5: Left: Fluid lensing reproduction tank concept. Right: setup used for outdoor seagrass dataset.*

### 3.1. Datasets

A number of datasets were created using either static images or videos. Image datasets were intended for models that made use of static images, like super resolution models. Video datasets were intended for sequence based models to test if they provide a better illustration of the effects of fluid lensing on an object over time.

#### 3.1.1. Checkerboard Dataset

The first dataset created was used to see if a model could re-create a static pattern that was influenced by fluid lensing – see Figure 6. The image was a simple checkerboard pattern and the dataset consisted of 397 images with water. Each of these images was compared to the same image without water.

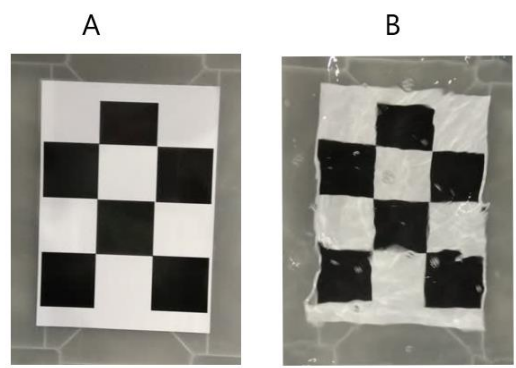

*Figure 6: Example images from checkerboard dataset. A) Target checkerboard image. B) Image with water influenced by fluid lensing.*

#### 3.1.2. Shapes Dataset

The second dataset consisted of randomly generated shape patterns that were created using the scikit-image python package[1]. Images consisted of between 3-10 shapes of different sizes. Shapes consisted of squares, rectangles, triangles, circles and ellipsis. Colours were randomised and shapes were allowed to overlap. Two sets of photos were taken of all the patterns. The first was images with no water – these were to be used as the target images. The second was images with disturbed water which were used for training and validation/testing. Thus each shape pattern had two images in the dataset – one with water and one without water. Images were then cropped so that only the shape pattern and tablet border were in the image. The border was retained for this dataset because the water caused the distorted shape pattern to appear above the border and removing this would remove important information. The images were then downscaled to 128x202 and 256x404. See Figure 7 for examples. Initially a dataset of 800 images of 800 different patterns was collected. This was later increased to 1200, the reasons for which are discussed later. 5% of the dataset is used for validation and 5% for testing. The camera used was a 16MP camera on a Nokia 6.1 smartphone.

---

[1] https://scikit-image.org/docs/dev/auto_examples/edges/plot_random_shapes.html

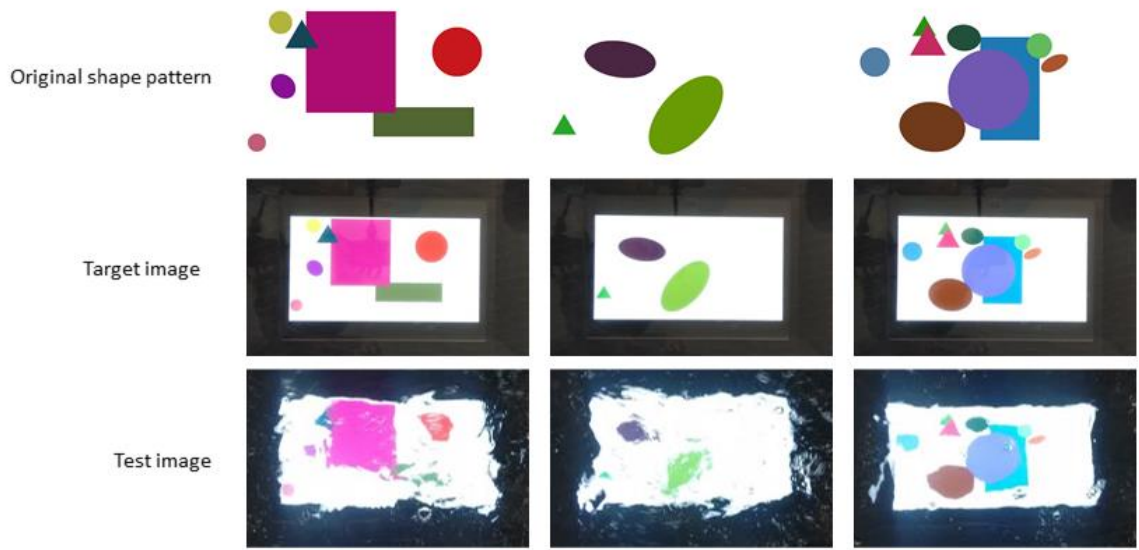

*Figure 7: Example images from shapes dataset.*

A third dataset was created following the same methodology except that videos were collected of the disturbed water images. 200 videos were made with each being 3 seconds at 27 frames per second. They were also cropped and downscaled to 128x202. The target was an image of the same shape pattern without water.

### 3.1.3. Outdoor Seagrass Dataset

The shapes dataset had a number of issues. The edge of the tablet is in all the images so models would always learn to add this edge back. This would impact potential test images that did not contain the same edge. The water was also too disturbed and often contained large bubbles on the surface. This does not reflect the actual conditions that seagrass monitoring data would be collected in and so overcomplicates the learning process for models.

The outdoor seagrass dataset was created to fix these issues and to also create a dataset with images that better reflect the environment where the model would be applied. This, however, presented some challenges. Because the objects that need to have fluid lensing effects removed are all under water an actual target image that does not contain fluid lensing effects cannot easily be collected.

Therefore, an intermediate step was devised that used images of outdoor objects that looked like objects that are found underwater. These images included 260 images of rocks, shells, sand, long grass and other plants sourced from Google image searches, 152 manually taken images of sand dune vegetation and seagrass washed up on Largs Beach, Adelaide, and 308 wheat crop photos taken by a drone during a field monitoring mission provided by Ken Clarke and Dillon Campbell (see Figure 8) This dataset totalled 720 images.

A similar process to the shapes dataset was used to recreate the fluid lensing effects. Videos of each image were taken using a Canon M50 at 50fps and were at 3-5 seconds long. Each video was cropped and downscaled to 256x172. The tablet border was also cropped out though this was difficult to do accurately because sometimes the fluid lensing distortions caused either the edge of the tablet to reappear or the edge of the image to disappear in some of the frames. The target image was the original image. Data was manually split to ensure that a variety of images were present in the validation and testing datasets. The training/validation/test dataset split was 630/45/45.

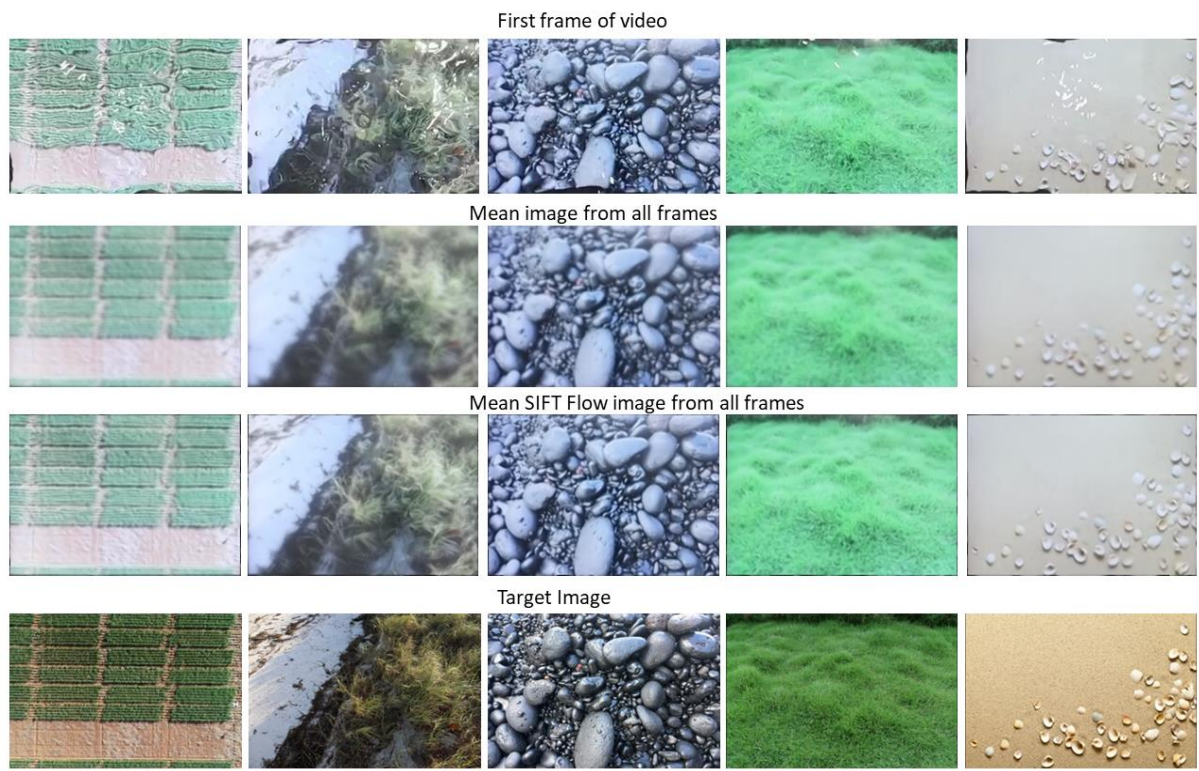

Figure 8: Example data from outdoor seagrass dataset.

### 3.1.3.1. Greyscale

A greyscale version of the outdoor seagrass dataset was created to see if the colour differences between the video and target data, discussed in 3.1.5, was impacting the final results.

### 3.1.3.2. SIFT Flow

SIFT Flow [6] is a method used to determine the displacement of points between frames even during variations in scale, illumination and transformation. It was also an important component in [3]. A variation of the outdoor seagrass dataset was created using SIFT Flow.

Firstly a mean image from n frames was generated for each video. Then, for each video frame, SIFT Flow descriptors are generated[2]. Using these descriptors, local matches are found between the frame and the mean image. These local matches are then warped onto the original frame to generate a SIFT Flow image. Using the SIFT Flow images, a mean SIFT Flow image is generated using 10, 25, 50, 100 and all SIFT Flow images. The resulting mean image appears to be slightly clearer and more stable than the mean image using the original frames – see Figure 8.

---

[2] The Python/Pytorch implementation of SIFT Flow was used for these steps. Code is available here: www.github.com/hmorimitsu/sift-flow-gpu.

### 3.1.3.3. Mean Image

A mean image generated from a video generally provides a more stable image than the individual frames, albeit a little blurry. Datasets of mean images using the mean of the first 10, 25, 50, 100 and all frames of each video was created.

### 3.1.3.4. Image Blurring

To test if this problem could instead be treated as an image deblurring problem a number of blurring methods were applied to the original dataset using a combination of methods:

- Averaging: Images are convolved with a box filter. The average of all pixels under the kernel area is computed and the central element is replaced.
- Gaussian Blurring: Similar to the above but instead of a box filter a gaussian kernel is used.
- Bilateral Filtering: Uses a gaussian filter like gaussian blurring but also incorporates pixel differences.

The blurring method was randomised during training. A random kernel of either 3, 5, 7 or 9 was selected to adjust the blur amount. For the bilateral filter a random sigma of either 0, 50, 100 or 200 was selected and for the gaussian filter a random standard deviation of either 0, 2, 4 or 6 was selected.

A number of deblurring models were tested which involved burring different images. This is discussed further below in 3.2.4.

### 3.1.3.5. Frame Rate Test Dataset

A small test dataset was created from 12 of the original test images to compare the model output when different frame rate video was input into a model that had only been trained using videos of one frame rate. Each test image was recorded using 25, 50 and 100 fps video. While care was taken to try and disturb the water a similar amount for each image it is impossible for the disturbance to be the same. Unfortunately the colour of the videos was different to the original videos so only visual evaluation can be performed.

### 3.1.4. Noarlunga Test Dataset

A small test dataset of underwater objects was collected from Noarlunga jetty. The conditions were wavier, the sun more reflective, the water depth greater, some underwater objects moved with the current and the camera height above the water greater than the model training dataset. 6 videos at 50 fps were taken. Even though the conditions were very different to what the models here are trained on and the conditions seagrass meadow monitoring is done, it was hoped that the output might still show if the model is learning anything useful that could be applied in real world conditions.

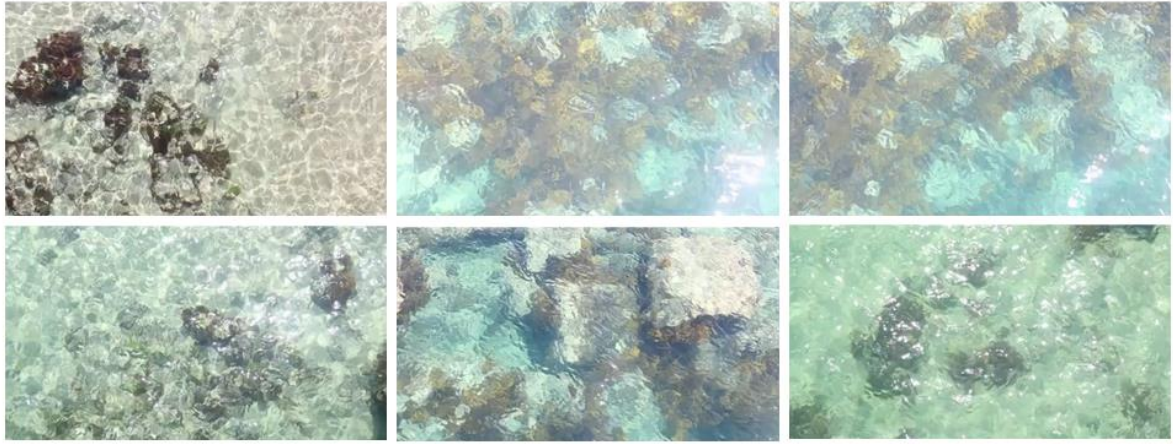

*Figure 9: First frame from Noarlunga test videos.*

### 3.1.5. Issues with data collection method

It can be seen in Figure 7 and Figure 8 that there is colour loss in the captured images. This is caused by taking images/videos of a screen – the further away or less zoomed in the camera is the greater the colour loss. This is an unfortunate side effect of this data collection method and adds additional complexity to the model as it needs to learn the colour mappings. It also complicates model evaluation and comparison if some models need to learn the colour mappings and others don't (eg: blurring hr images). While this is tolerable for a proof of concept it would not be suitable for real world applications as the colour mappings would be incorrectly applied to real world data.

## 3.2. Models

A number of models were trained and evaluated on the fluid lensing datasets.

### 3.2.1. EDSR

EDSR [7] is a model originally designed for super-resolution tasks where a low resolution image could be converted to a high resolution image. Such models are being experimented with for other similar tasks such as the removal of cloud from satellite photos. EDSR was implemented to see if it could also be effective at removing fluid lensing effects.

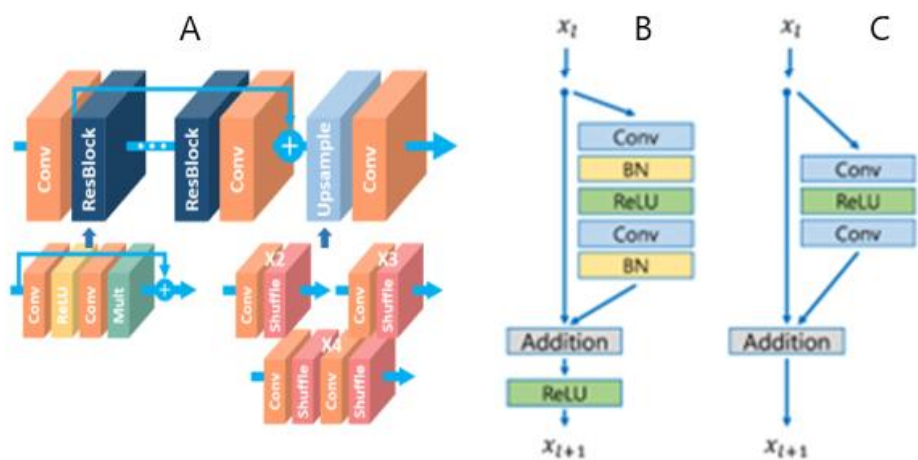

*Figure 10: A) EDSR model structure. B) Original Resnet residual block. C) EDSR residual block. [7]*

The model used is based on the convolutional Resnet [7] architecture where a series of residual blocks containing convolutional layers are stacked together. However, EDSR uses modified residual blocks that don't use batch normalisation or block output activation layers. This model was originally trained by passing patches of low resolution images through the network, upscaling the image and then comparing the patch to the high resolution equivalent. The model contains a number of hyper parameters that can be configured including the number of residual blocks, the number of features in the convolutional layers and output scale. Images are normalised using the mean RGB values of the target dataset.

The training process is slightly modified for training the network to remove fluid lensing effects from the shapes image dataset. Images with disturbed water are passed through the network with the output image being compared to the same shape pattern image without water.

### 3.2.2. Convolutional LSTM

The second model tested was a convolutional LSTM [8] and auto encoder variant. LSTMs [9] were originally designed for sequenced based tasks where data is input into a model one at a time and in sequence. By incorporating long and short term memory the model is able to make decisions around which information is useful and should be kept and which information should be discarded. This allows it to learn complex relationships between sequences. This was achieved using a series of gates that incorporated fully connected layers. Using fully connected layers, however, makes the model unsuitable for handling spatial data as input data needs to be flattened resulting in the loss of spatial information.

Convolutional LSTMs were designed to support spatial data and spatial forecasting. By replacing the fully connected layers with convolutional layers spatial information can be input and transferred between states. Like LSTMs, Convolutional LSTM can also be stacked to increase the depth of the network and used in an encoder-decoder model.

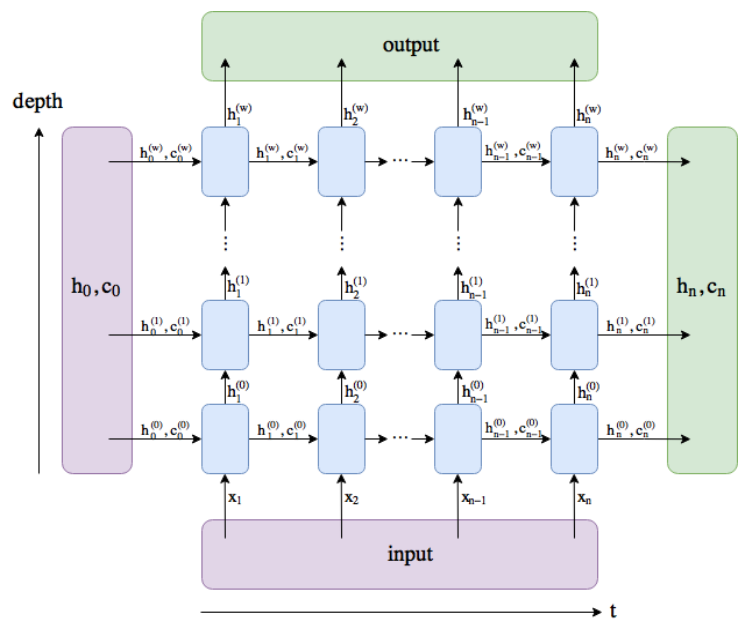

*Figure 11: LSTM flow diagram with hidden state h, cell state c, input of sequence x and depth of w. Output size, depth and input size are configured by the hyper parameters. [10]*

Convolutional LSTMs have been used for tasks including video next frame predictions, activity recognition and weather forecasting. Here it will be used to try to remove the effects of fluid lensing. This model is trained by passing in a sequence of frames of water disturbed images and comparing the output to images without water.

Hyper parameters for convolutional LSTMs and auto-encoder consist of the number of layers (stacked LSTMs), the number of features and kernel size in the convolutional layers and the length of the input sequence.

### 3.2.3. Spatial Temporal Convolutional Network (STCN)

Temporal convolutional networks (TCNs) [11] were designed to capture long range patterns for action segmentation and detection using a hierarchy of convolutional filters. Using a sequence of stacked 1D inputs, computations are performed across all layers simultaneously. This means that each time step is updated simultaneously instead of sequentially per frame and convolutions are computed across time. This is achieved through casual 1D dilated convolutions. As data progresses through the network, the dilation factor increases producing a larger receptive field. TCNs have proved to perform better than residual networks (LSTM, GRU) across a number of tasks. As TCNs use 1D inputs it is less suited to handling spatial data and image restoration tasks in its original form. Therefore, the key concepts were taken and adapted to support spatial data, hence STCN.

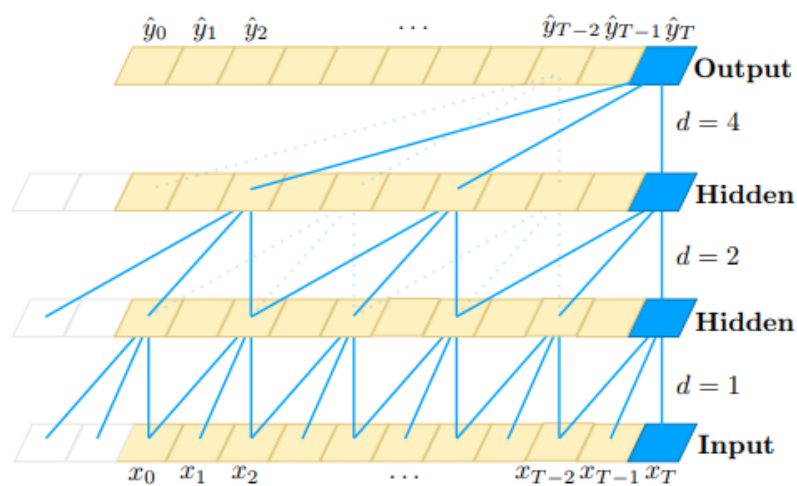

*Figure 12: Dilated convolutions in a TCN with dilation factor d, kernel size of 3, input x and output y [12].*

Two variations were developed. The first variant, stacked STCN, makes use of 2D convolutions. The input sequence is stacked along the channel axis such that the input shape is *height * width * depth* where *depth = image channels * sequence length*. The kernel size is *3 * 3 * depth*. The second variant, non-stacked STCN, used 3D convolutions resulting in an input shape of *sequence length * height * width * channels* and uses a kernel of size 3.

The rest of the network structure is the same for both variants. First the input is normalised by subtracting mean RGB values from the target dataset. The main network contains a number of layers *L* each of which contained a number of temporal blocks *B* with the same layers as C in Figure 10. Each block uses a dilation factor $d = 2^{BN-1}$ where *BN* reflects the block position in its layer. Thus block 3 in layer 1 has the same dilation factor as block 3 in layer 2. As *d* increases the amount of padding also increases which increases the size of the convolutional layer and the GPU memory required. This can restrict the number of blocks that can be placed in a layer. The final network layer is a convolutional layer with an output of *size = image channels* so that the network produces a single output image. The mean RGB values from the target dataset is then added to this output. The convolutional layers have 32 filters when processing a sequence of frames, as larger values tended to result in unstable training. To further stabilise training residual scaling [13] with a factor of 0.1 is used. Including batch normalisation and dropout layers in the temporal block was also tested but model performance was noticeably reduced.

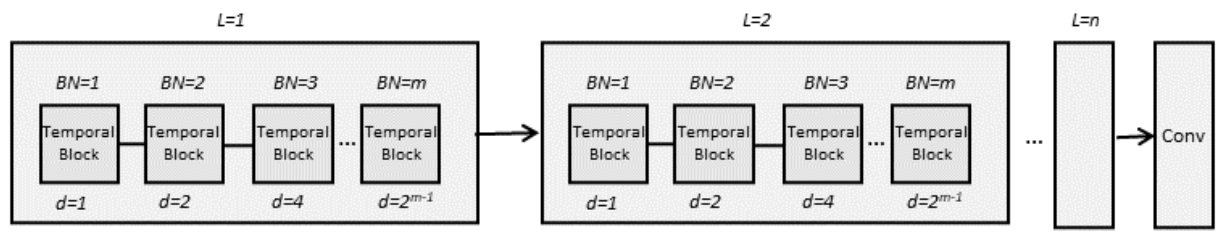

*Figure 13: STCN architecture.*

This model can be trained by either by passing in a sequence of video frames or a single image. When using a single image as input the model is comparable to EDSR with the addition of dilation. Processing single images also allowed a larger number of filters in the convolutional layers.

### 3.2.4. Image Deblurring Models

*3.2.4.1. Mean images*

Using STCN, the mean image from 3.1.3.3 is used as the input image with the output being compared to the matching hr image. The aim of this model was to train the network to remove the blur from the mean image.

*3.2.4.2. Mean SIFT Flow images*

Using STCN, the mean SIFT Flow image from 3.1.3.2 is used as the input image with the output being compared to the matching hr image. The aim of this model was to train the network to remove the blur from the mean SIFT Flow image.

*3.2.4.3. Blurring video frames*

During training, one of the blurring methods mentioned in 3.1.3.4 is used on all the frames of a single video. These frames are used as single input images to STCN and compared to the original frame. During testing the mean image is passed into the network with the output being compared to the original hr image.

This model is intentionally made to overfit on frames from one video meaning that a model needs to be trained for each video. The aim of this model was to train the network to remove the blur from the frames so that the blur could be removed from the mean image.

*3.2.4.4. Blurring hr images*

The original hr images are blurred using one of the blurring methods mentioned in 3.1.3.4 and are used to train the STCN model. For testing, the mean image is passed into the network with the output being compared to the original hr image. The aim of this model was to train the network to remove the blur from the hr images so that the blur could be removed from the mean image.

### 3.2.5. Model Evaluation

Models are evaluated on loss, peak signal to noise ratio (PSNR) and visual comparison. For loss, mean square error, root mean square error and L1 loss functions were tested with L1 loss being the primary loss function used. These are all pixel-wise loss functions that are commonly used in image based models. PSNR is commonly used to measure the quality of images that undergo reconstruction, for example, low resolution to high resolution. The signal is the original image and the noise is the error present in the comparison image. As mentioned in 3.1.5, using loss and PSNR wont always be a reliable method of evaluation so visually evaluating the model output is also heavily relied on.

For optimisation, the Adam optimiser was used. The Asmgrad algorithm for the Adam optimiser was used as it helped to stabilise learning. Gradient clipping was performed if gradients grew larger than 1. Models were train and tested on Nvidia V100 GPUs on the AIML Deep purple cluster.

Training data used in the models underwent data augmentation to randomly flip the images/frames either horizontally and/or vertically during training. For the videos, frame sequences are taken from a random start position in the video and are also randomly reversed.

### 3.2.6. Model Electricity Consumption

The training of machine learning models is requiring an increasing amount of electricity [14]. It was hoped that the total electricity usage of this project could be calculated using a similar approach to [14] but unfortunately AIML does not keep any usage logs that could be used to calculate this.

## 4. Results and analysis – EDSR and Convolutional LSTM

### 4.1. Checkerboard dataset on EDSR

Checkerboard images that were influenced by fluid lensing were successfully recreated using EDSR. However, as the model was only learning to produce a single static image the model is of little actual value.

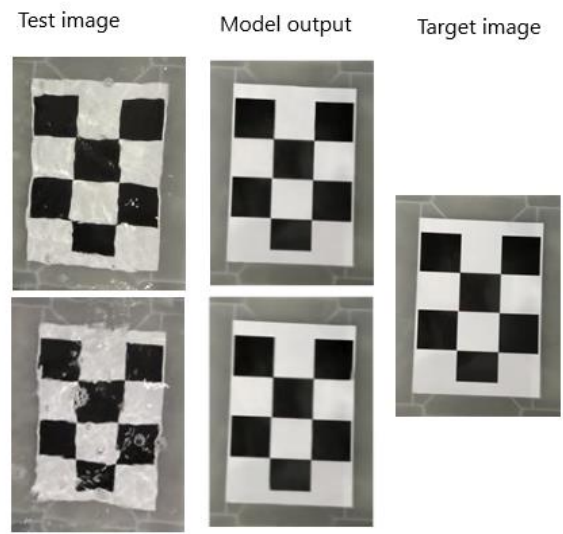

*Figure 14: Example test output from checkerboard EDSR model.*

### 4.2. Shapes image dataset on EDSR

After experimenting with EDSR using the initial 800 image shapes dataset, a similar pattern emerged from the results. Many of the models appeared to be overfitting (see Figure 15). The training loss continues to decreases but the validation loss flattens at around 10. Similarly, the PSNR for training continues to increases but the validation PSNR flattens at around 19.5.

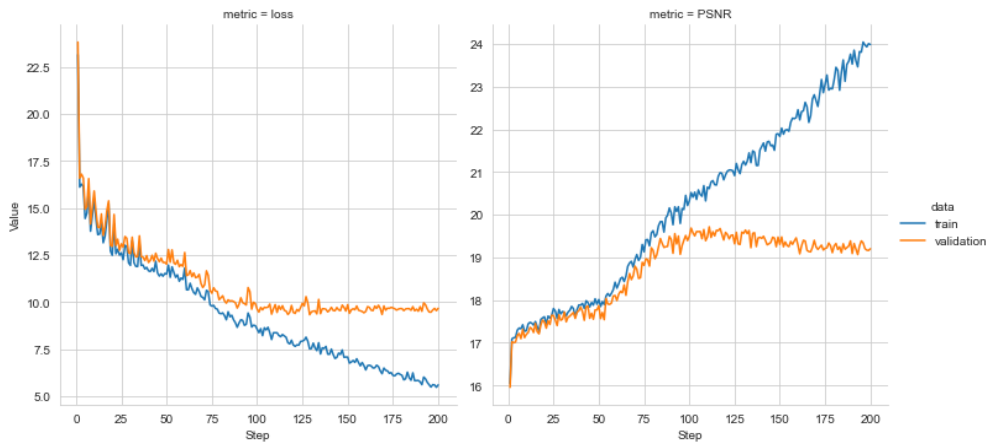

*Figure 15: L1 loss and PSNR results for EDSR model with 18 residual blocks, 256 convolutional features Adam optimiser epsilon of 1e-5 using amsgrad. Model trained on the 800 image shapes dataset.*

To attempt to rectify the overfitting the dataset was expanded from 800 images to 1200. Unfortunately, very similar results were observed, shown in Figure 16, with the training and validation curves splitting after reaching the same loss and PSNR values as the smaller dataset.

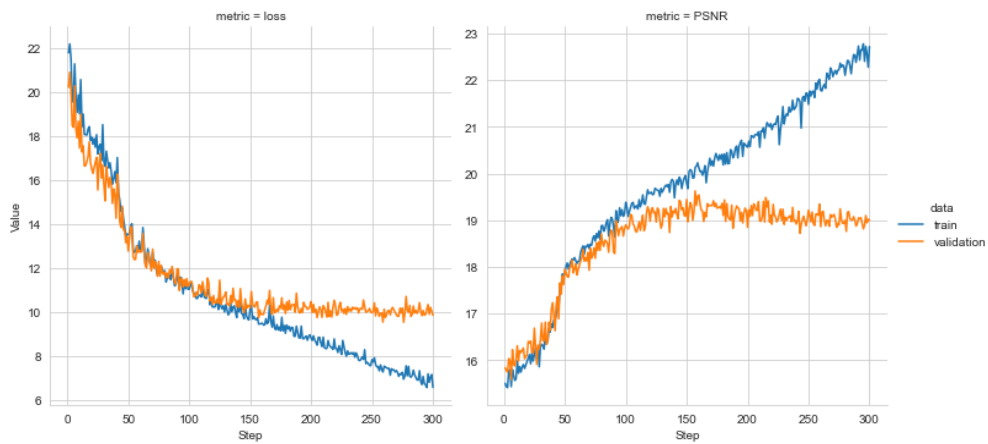

*Figure 16: L1 loss and PSNR results for EDSR model with 32 residual blocks, 256 convolutional features Adam optimiser epsilon of 1e-4 using amsgrad. Model trained on the 1200 image shapes dataset.*

This suggested a different conclusion: using single images does not allow the model to learn enough information about fluid lensing to remove its effects. This conclusion is supported by the images in Figure 18. For example. the green rectangle in the first target image is so distorted in the test image that the output for that image attempts to create two shapes from the distortions. To a human eye, this seems to be a logical decision as the distorted image contains two green sections. These results led to the decision to test sequence based models using videos.

### 4.3. Shapes videos on Convolutional LSTM and autoencoder

Extensive testing was performed on both convolutional LSTMs and autoencoders using combinations of image normalisation, model outputs and model hyper-parameters. Very few models showed any ability to learn information with training loss either remaining fairly constant throughout training or increasing when more complex models were tested.

The few that did show signs of learning didn't learn much and still performed poorly on the validation data. The better performing models were auto-encoders but when more complex variants were tested or when a sequence length greater than 5 was used these models also performed poorly. Figure 17 shows a better performing model. While it looks like this model is overfitting, as the validation metrics only improve for the first 2-3 epochs, because so few models actually show signs of learning it seems more likely that convolutional LSTMs/autoencoders in their current form are not suited to removing the effects of fluid lensing.

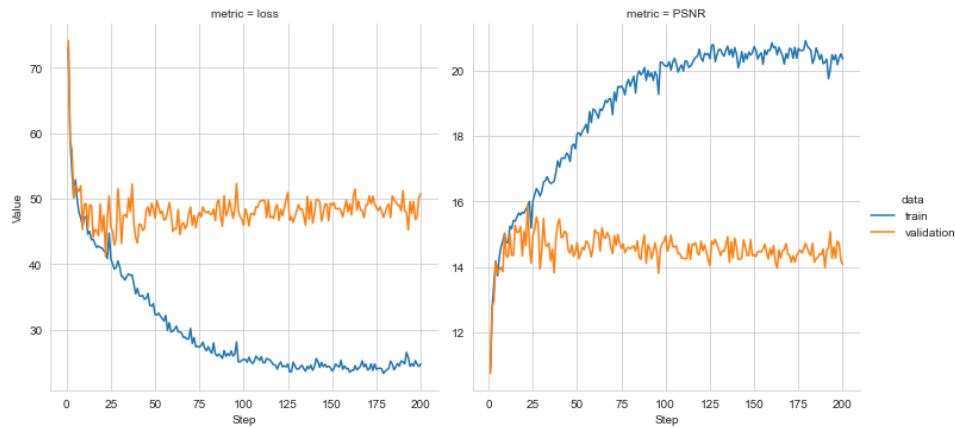

*Figure 17: Better performing LSTM autoencoder containing two layers with RMSE loss, hidden dimensions of [128, 64], kernel size of 3 and sequence length of 5. Sequence frames were normalised at the start using the average sequence image and the mean comparison data set RGB values were added at the end.*

It is clear from the output images in Figure 18 that the shapes are not close to being reconstructed. While the training images are marginally better it is still not clear what the shapes in the image should be. It seems that either LSTMs are not well suited to this task or more information needs to be provided to support the learning process.

### 4.4. Shapes videos on non-stacked STCN

STCN performs fairly well on the shapes videos dataset. It manages to reconstruct the larger shapes reasonably well and even manages to remove bubbles for some images but still struggles with the smaller shapes. Because of the issues with the shapes dataset mentioned in 3.1.3 only limited testing of STCN on the shapes dataset was conducted. A more through analysis of STCN is performed on the outdoor seagrass dataset. Results for STCN on the shapes dataset are shown in Figure 18.

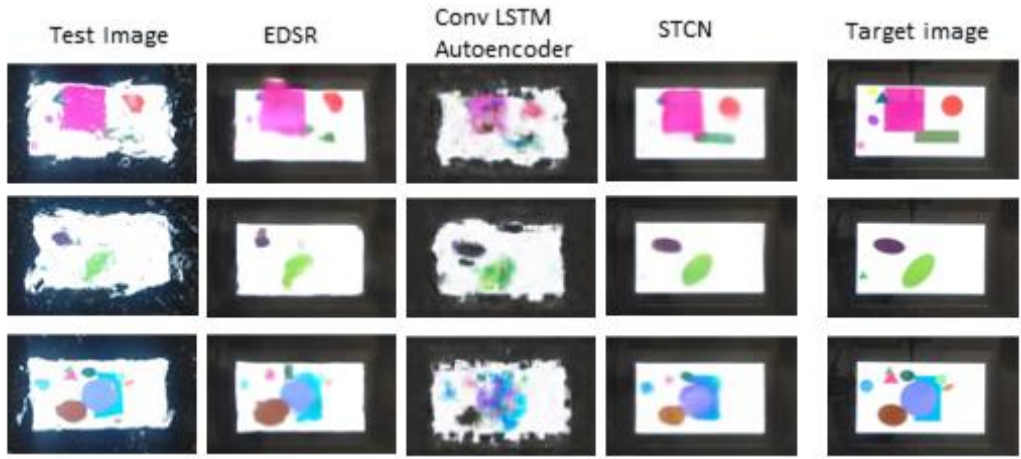

*Figure 18: Model test output on the shapes datasets. The images for EDSR and Conv LSTM models were generated using the models mentioned in Figure 16 and Figure 17. STCN used a non-stacked input of 25 frames, 7 layers and 4 blocks.*

## 5. Results and Analysis – STCN

To further investigate STCN and its suitability to this task, the performance between different hyperparameters is compared using the outdoor seagrass dataset. Figure 19 shows the performance differences between models trained on different sequence length inputs for both STCN variants. It is clear that increasing the sequence length improves the model evaluation metrics. It is also clear that the non-stacked STCN variant outperforms the stacked STCN variant as the sequence length increases. The importance of dilation is also very noticeable. The model with no dilation (green) that had an input sequence length of 25 performed noticeably worse than equivalent models that used dilation.

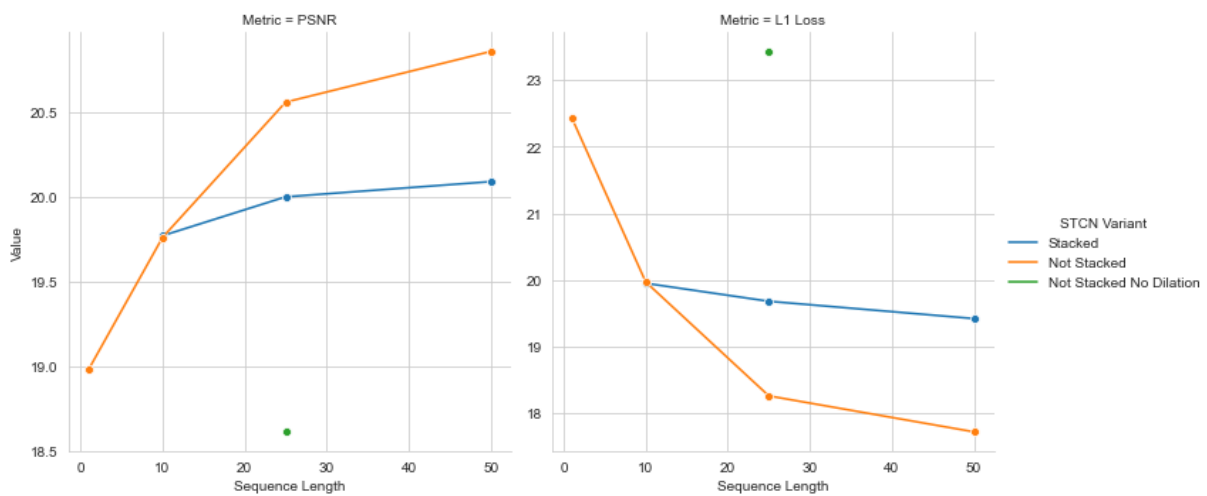

*Figure 19: Test dataset result comparisons between different sequence length inputs for the two STCN variants. All models used 7 layers * 4 blocks. Each model was trained for 200 epochs. The stacked variant was not specifically tested using a sequence length of 1 because the model would be identical to the non-stacked variant with a sequence length of 1.*

The model output can be used to further compare the differences when increasing sequence length. In Figure 20 it can be seen that for the non stacked variant, the clarity of the image increases as the sequence length increases. The individual rocks are better defined and the gaps between the crop sections become clearer. This is also observable in the stacked variant but the changes are less dramatic and the final images are not as clear as the non-stacked variant.

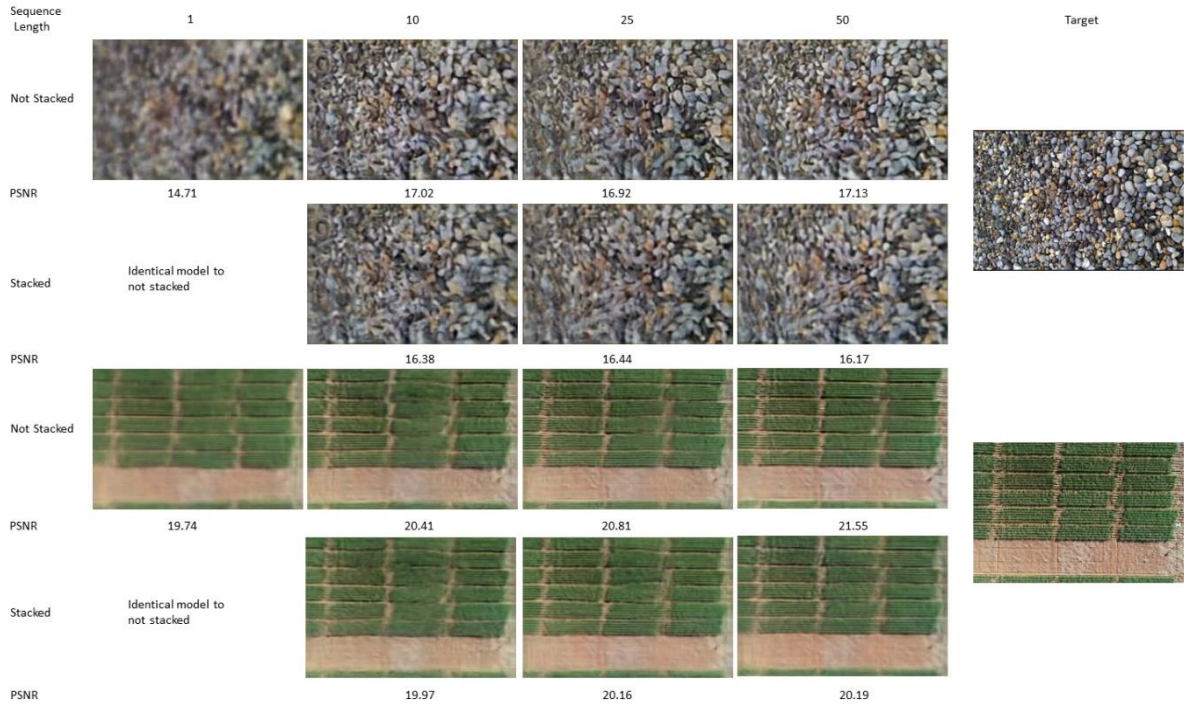

*Figure 20: Comparison of test dataset output from models using different sequence lengths in Figure 19.*

To investigate how the structure of the network impacted model performance a number of models with a similar number of total blocks (14, 15 or 16) were trained and compared. These blocks were spread across a different number of layers. All models had an input sequence length of 10 frames. As the non-stacked variant of STCN performed better than the stacked variant in the sequence length tests, only the non-stacked variant was tested. Table 1 shows the evaluation metrics for these models.

These results again show the importance of using dilated convolutions. Models with a lower number of blocks per layer have noticeably lower performance. Performance peaks for models with 5 blocks and then recedes for models with larger number of blocks. Models with a higher block count than those listed would have required significantly more GPU memory to test and considering the size of the input frames, the resulting receptive field would have covered more than half of the image. Therefore, models with more than 7 blocks were not tested.

| Model (Layers * Blocks) | PSNR | L1 Loss |
| --- | --- | --- |
| 2 * 7 | 19.65 | 20.30 |
| **3 * 5** | **19.89** | **19.82** |
| 4 * 4 | 19.68 | 20.17 |
| 5 * 3 | 19.38 | 20.95 |
| 7 * 2 | 18.81 | 22.04 |
| 15 * 1 | 18.97 | 21.87 |

*Table 1: Test dataset result comparisons between different layer and block combinations. Each model used a sequence length of 10 frames and was trained for 100 epochs.*

To investigate if the colour issues mentioned in 3.1.5 were affecting model performance, the output from two identical models are compared, shown in Figure 21. One model was trained using the RGB dataset and the other was trained using the greyscale dataset. There seems to be no obvious improvement in model output when using the greyscale images. It is also more difficult to determine the boundaries of some objects in the greyscale images. The evaluation metrics could not be used for this comparison because when the metrics are calculated for the greyscale images they are only calculated for 1 channel in comparison to 3 channels for the RGB images.

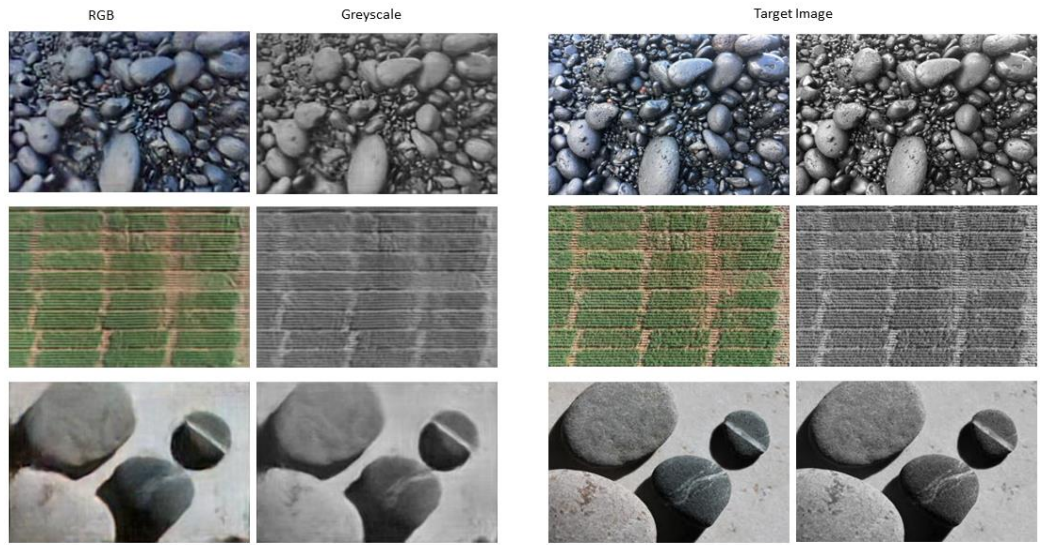

*Figure 21: Test data output comparison between using RGB and greyscale images. Both models were trained for 200 epochs and had 7 layers * 4 blocks.*

The best performing STCN model was a non-stacked variant using 7 layers with 5 blocks, a sequence length of 50 and trained for 200 epochs across 4 days and 16 hours using 3 GPUs. This model produced a test dataset PSNR value of 21.65, 0.79 higher than the best performing model in Figure 19. Some model test output images are shown in Figure 26.

While the model is quite successful in recreating larger objects, the finer details of these objects are more challenging to recreate. This can be seen in the 4$^{th}$ and 6$^{th}$ image. The outline of the objects is reasonable but the objects themselves are still a little blurry. Importantly though, the fluid lensing effects are largely removed from the images– there is no light reflection and the distortions caused by the water are largely removed. The edges of the images are the least clear sections, probably because of the challenges with cropping the video as mentioned in 3.1.3. Its also possible this model could be improved slightly with more training as the training and validation metrics were still improving gradually.

The frame rate test dataset was run on the best performing STCN model above. Some example output is shown in Figure 22. It is difficult to draw reliable conclusions from this however because of the challenges in creating such a comparison dataset. The 100 fps images seem more affected on the boarder by the tablet distortions, probably because they appear in more of the frames. For the pebble image the 25 fps is actually clearer than the other images. More testing is needed to determine the actual impact of using different video frame rates for both training and testing.

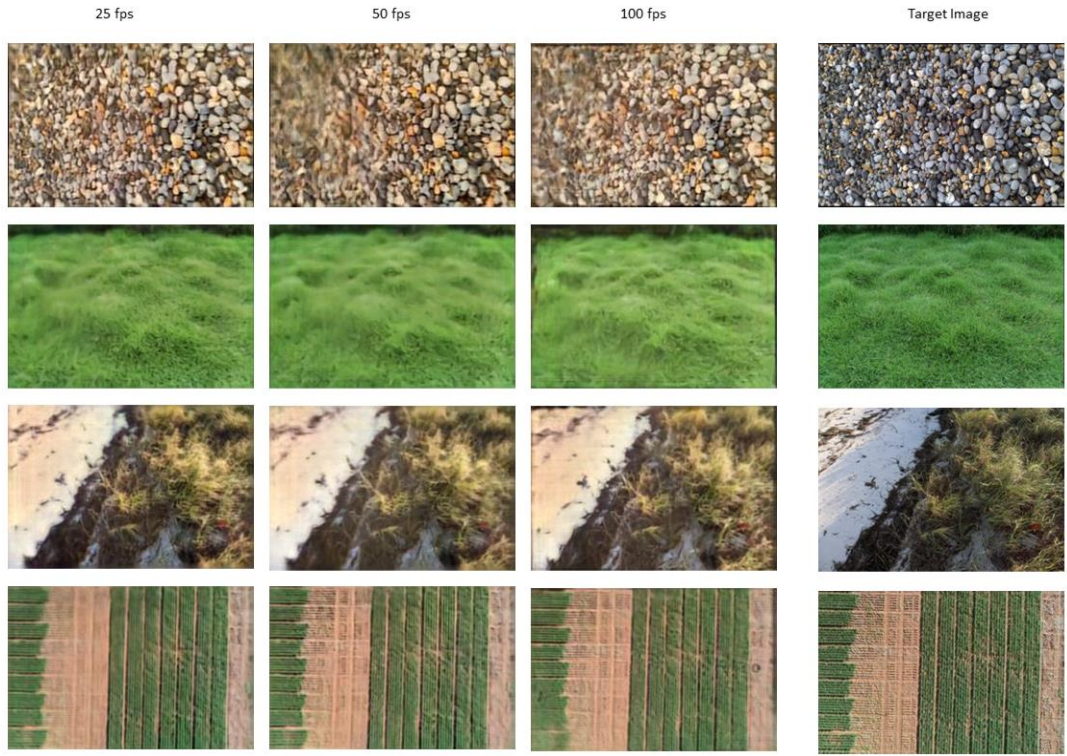

*Figure 22: Test output when testing with different frame rate videos on STCN 50 frames model.*

Some sequence based models produce useable output when different length sequences are input into the model. To see if this was also the case for STCN the test dataset was passed into the best performing STCN model using different sequence lengths. Output is shown in Figure 23. It is clear that inputting a different length sequence into the model than it was trained on does not provide very useful output. It would be interesting to try training a model using a random sequence length for each batch and re-run this experiment to see if the results are different.

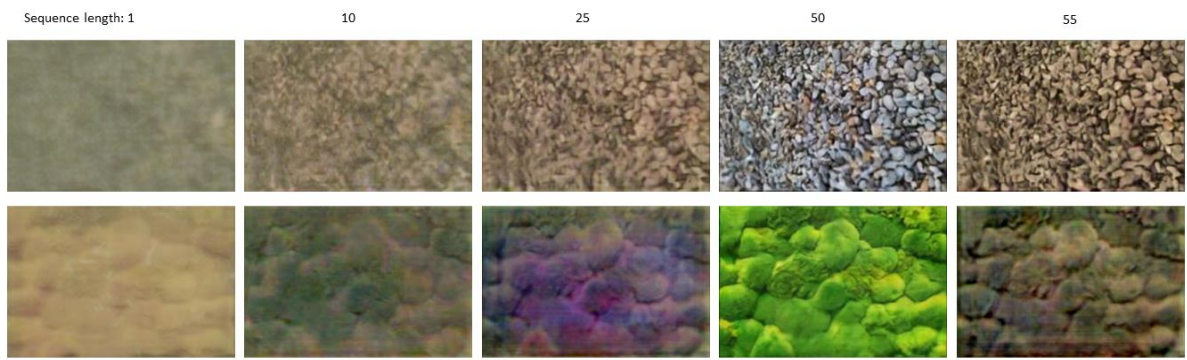

*Figure 23: Model test output when passing input of different sequence lengths into a trained STCN model. Figure 22 shows the target image for the images in this figure.*

## 6. Results and Analysis – Deblurring

Models here were trained using single images on the STCN model using 5 blocks and 7 layers with the convolutional layers having 128 filters.

## Image Blurring

Example test output for the deblurring models is shown in Figure 26 (mean image, blurring hr images and mean SIFT Flow image) and Figure 24 (frame blurring). The frame blurring method is the least successful method. It seems that instead of learning to remove the image blurring the model is learning to re-create the fluid lensing effects. The resulting test output image is less clear than the original input frames. The hr blurring model images seem to retain more water in the image – this can be seen in the 4$^{th}$ image, though the image colouring issues makes this model difficult to compare to the other deblurring models. To properly compare the hr blurring model a training dataset that did not require the model to learn colour mappings would be needed.

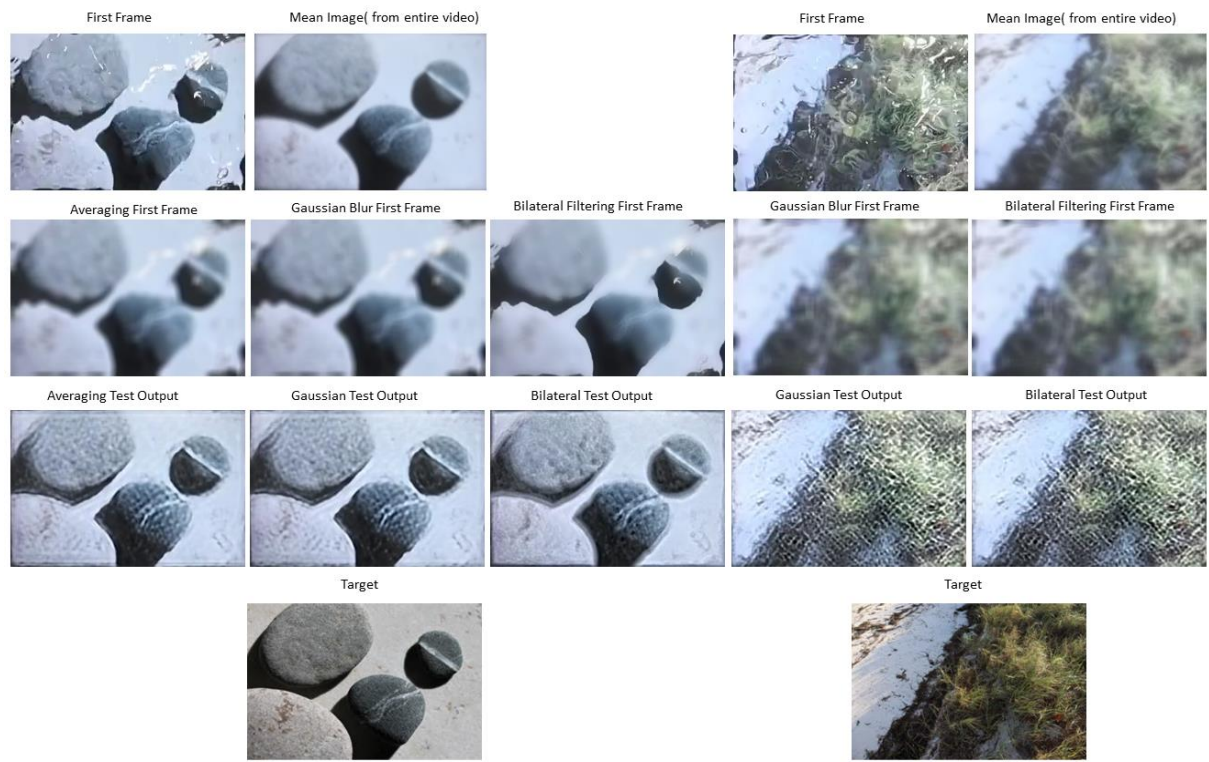

*Figure 24: Example training data and model test output for frame blurring model.*

## SIFT Flow and Mean Image

The SIFT Flow and mean image models create noticeably clearer images than all other models tested. Figure 25 and Table 2 show that increasing the number of frames to create the mean image steadily improves the evaluation metrics and the clarity of the output image – the finer details of the image show quite well and the edges of the image are clearer. This could be due to the water in the later frames in the video being much more stable thus allowing for a clearer mean image when all frames are used. There are, however, a small number of cases where using only 50 or 100 frames produce images with a higher PSNR than when using all images (6 out of 46 test images for SIFT Flow).

Using SIFT Flow for data pre-processing provides a small improvement overall in comparison to just using the mean image. This improvement is consistent across all models trained (Table 2). This improvement is not consistent between all test images – for some images just using the mean image produces better results. Further testing on real world data may be needed to determine if using SIFT Flow improves model performance or if using 100+ frames to generate the mean image produces a similar performance improvement.

| Dataset | No. Frames | PSNR | L1 Loss |
|---|---|---|---|
| SIFT Flow Mean Image | 10 | 19.65 | 20.18 |
| | 25 | 20.43 | 18.49 |
| | 50 | 20.86 | 17.79 |
| | 100 | 21.61 | 16.22 |
| | All Frames | **22.85** | **13.81** |
| Mean Image | 10 | 19.56 | 20.45 |
| | 25 | 20.33 | 18.79 |
| | 50 | 20.78 | 17.97 |
| | 100 | 21.69 | 15.99 |
| | All Frames | 22.77 | 14.06 |

*Table 2: SIFT Flow mean image and mean image test results for images made from a different number of frames.*

These models also produces much clearer overall results overall than all other models tested and could possibly be improved with more training as the evaluation metrics are still improving at 200 epochs. Model training also requires much less computational power and time than STCN – completing 200 epochs on 1 GPU takes around 22 hours. It is worth noting however that when using the same number of frames STCN with 50 frames does produce better results than the SIFT Flow and mean image models with 50 frames.

| Model | L1 Loss | PSNR |
|---|---|---|
| Mean Image | 14.06 | 22.77 |
| Mean SIFT Flow | **13.81** | **22.85** |
| STCN 50 Frames | 16.08 | 21.65 |

*Table 3: Test dataset evaluation metrics for final deblurring and STCN models. The HR Image blurring model is not included as the model did not learn the colour mappings - this impacts the metric comparison.*

| No. Frames | 10 | 50 | 100 | All | Target |
|---|---|---|---|---|---|
| SIFT Flow mean Image Model | | | | | |
| PSNR | 20.17 | 23.17 | 24.46 | 26.27 | |
| Mean Image Model | | | | | |
| PSNR | 20.08 | 22.86 | 23.11 | 27.25 | |
| SIFT Flow mean Image Model | | | | | |
| PSNR | 15.39 | 16.51 | 17.68 | 18.40 | |
| Mean Image Model | | | | | |
| PSNR | 15.56 | 16.89 | 17.84 | 18.59 | |
| SIFT Flow mean Image Model | | | | | |
| PSNR | 19.58 | 21.10 | 22.28 | 23.73 | |
| Mean Image Model | | | | | |
| PSNR | 19.72 | 21.38 | 22.26 | 23.55 | |
| SIFT Flow mean Image Model | | | | | |
| PSNR | 23.15 | 24.51 | 24.86 | 23.89 | |
| Mean Image Model | | | | | |
| PSNR | 23.07 | 23.23 | 24.87 | 23.57 | |

*Figure 25: SIFT Flow and mean image model test results for datasets created using a different number of frames.*

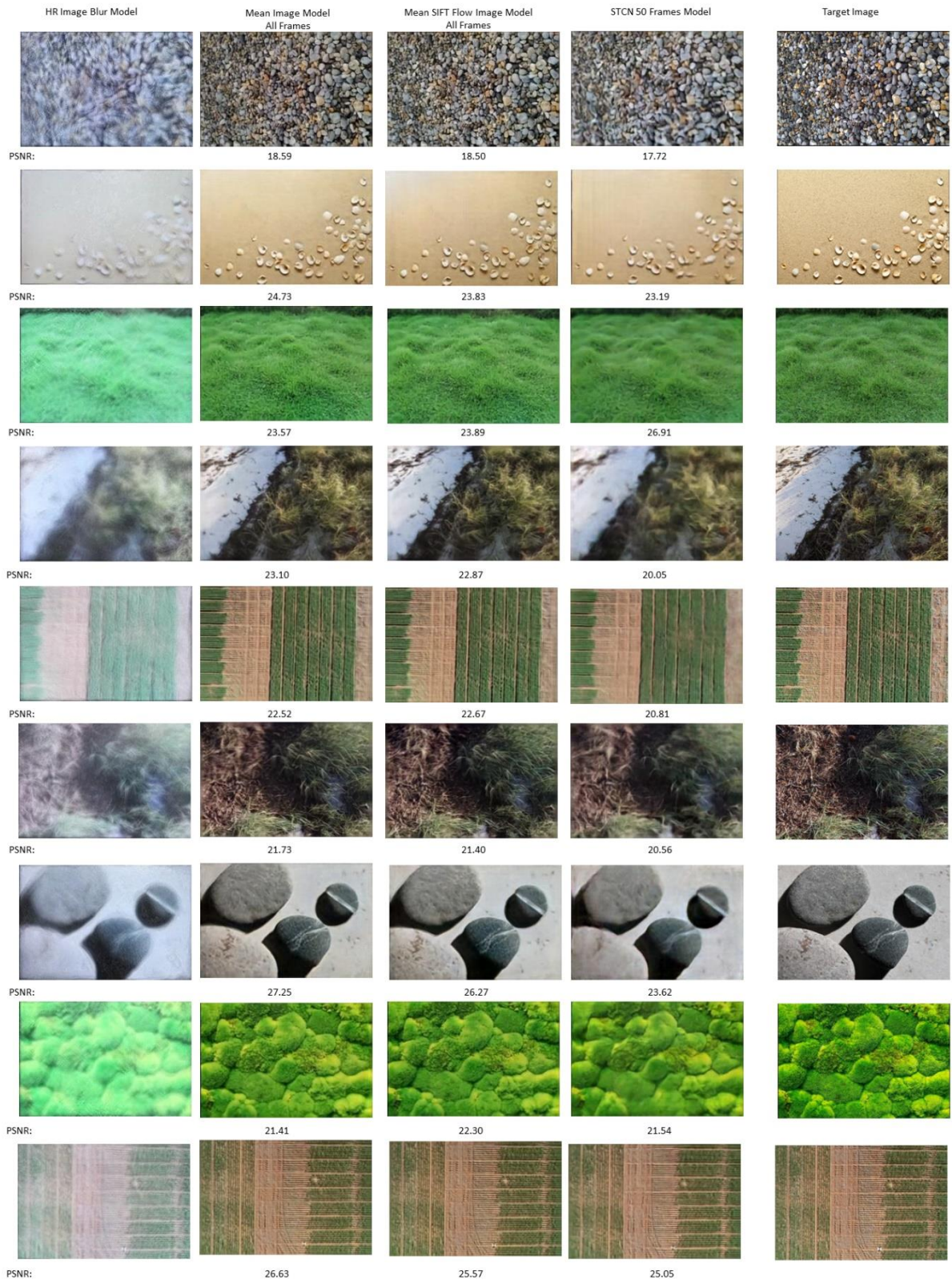

*Figure 26: Test dataset output comparison for STCN and the deblurring models. All models used a 7 layer * 5 block STCN model. The deblurring models using a sequence length of 1 and 128 convolutional filters. PSNR was not calculated for the HR image blurring model because the model did not learn the colour mappings..*

# 7. Testing on real world data – Noarlunga dataset.

The videos from Noarlunga were tested on the best performing STCN and SIFT Flow models from Sections 5 and 6. The model output images are shown in Figure 27. Considering the significant differences in conditions mentioned in 3.1.4 the STCN model appears to do an OK job at removing the effects of fluid lensing on shallower objects. Most of the surface sun reflection is removed and the distortive effects of the waves are reduced. The underwater bands of light are still present in the output images but these effects were poorly reflected in the training data so this is not surprising. The output is also clearer than the mean image. The SIFT Flow model does not perform as well here with only a little improvement over the mean image. It is clear though that both models struggle with objects in deeper water. Objects in shallower water (1$^{st}$ image, for example) are much clearer than those in deeper water (2$^{nd}$ and 4$^{th}$ images, for example). These results are still encouraging for future work given the conditions the videos were taken in.

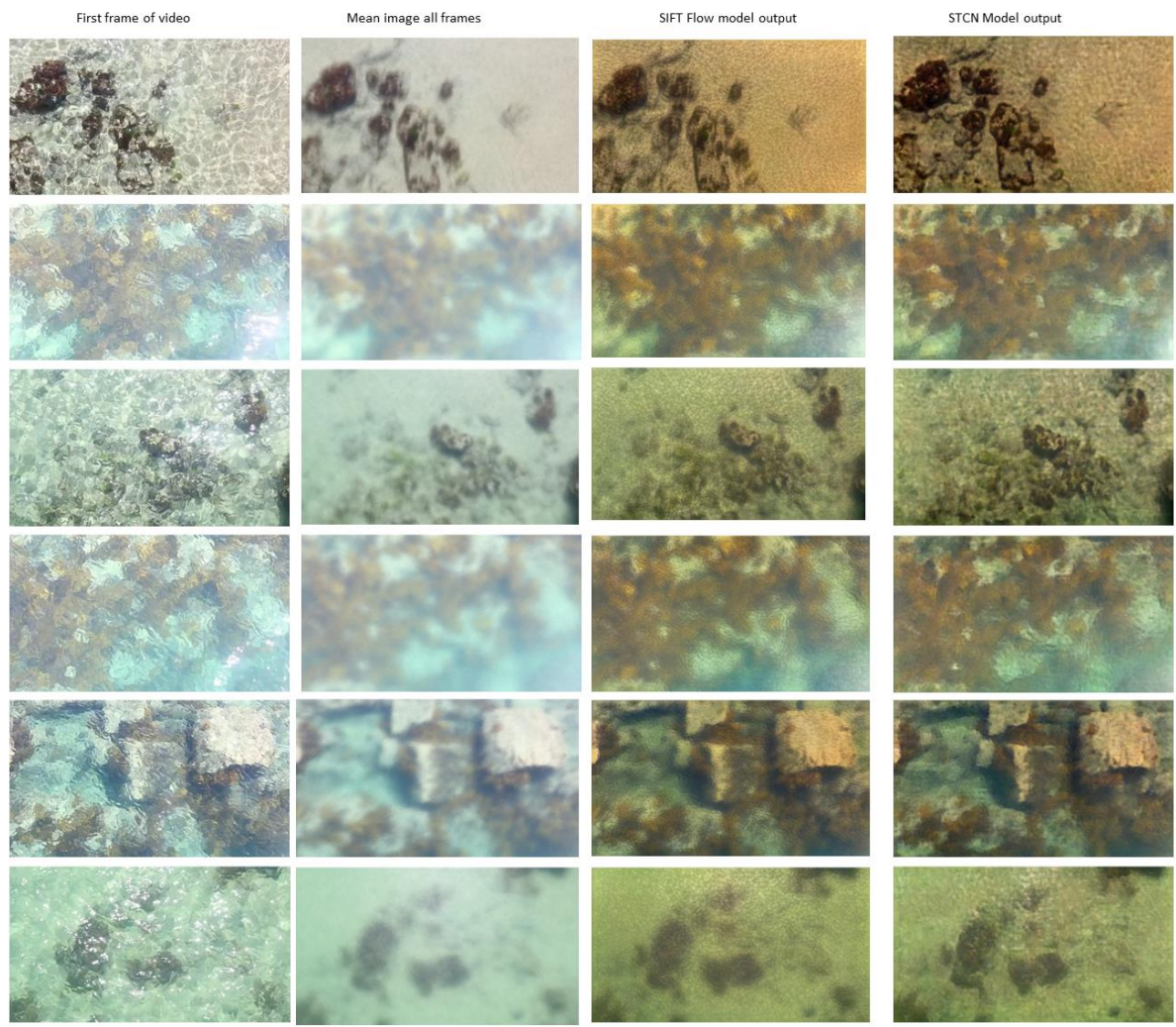

*Figure 27: Noarlunga test dataset model output. Unfortunately the output of the models is affected by the colour mappings learnt by the model.*

## 8. Future Work

The challenge with using a machine learning approach to this problem is scalability. Creating a dataset that could be used for a real world application would require data to be collected:

- at different water depths up to 10m-12m
- with the sun at different positions in the sky relative to both the water and the position of the camera
- at heights similar to how it would be collected in the field
- with enough variety of underwater objects
- with objects (like seagrass and seaweed) moving with the current

Creating such a dataset could possibly be done in an outdoor pool by placing objects (sand, rocks, seagrass like plants etc) on the bottom of an empty pool to create target data and gradually filling it up with water to create training and testing data. To get enough variety in the images this process may have to be repeated. However, this process would still face issues similar to the datasets created here. When cropping the pool out of the photos or splitting the pool photo into smaller photos the fluid lensing distortions along the edge of the images may still include unwanted information or remove important information though this would happen when taking videos in the real world too.

It is also difficult to evaluate real world data for the same reason it is hard to create a real world dataset – everything is already underwater. Manual evaluation through underwater photography would probably be required to validate model output. Nevertheless, the next step in evaluating this approach to removing fluid lensing effects from images would be to create a dataset that meets the above requirements and train and evaluate a model using this and real world data.

Some other future work could include:

- Comparing model output to the output generated by the NASA fluid lensing algorithm.
- Incorporating model attention layers has improved model performance for a number of tasks. Including model attention mechanisms for STCN or deblurring models could be tested to see if it improves performance.
- Compare models trained using different frame rate videos

## 9. Conclusion

A number of datasets were created through the replication of fluid lensing in a test tank. These datasets were tested on EDSR, convolutional LSTM, STCN and a variety of deblurring models. It is shown that STCN with mean SIFT Flow images was the most capable model tested for removing the effects of fluid lensing. This model was able to produce a reasonably clear image which, if the results are scalable, could be useful in supporting the monitoring of seagrass regrowth. STCN with SIFT Flow should be a good starting point for testing a dataset that reflects real world conditions.

## 10. Thanks

Thanks to Tat-Jun Chin (Australian Institute of Machine Learning) and Kenneth Clark (University of Adelaide) for supervising this research. TJs knowledge in machine learning and Kens knowledge of remote sensing was invaluable.